\documentclass[twocolumn,english,aps,superscriptaddress, pra,groupedaddress]{revtex4-2}

\usepackage[latin9]{inputenc}
\usepackage{amsmath}
\usepackage{amssymb}
\usepackage{graphicx}
\usepackage{babel}
\usepackage{mathrsfs}
\usepackage{amsfonts}
\usepackage{epstopdf}
\usepackage{caption}
\usepackage{multirow}
\usepackage{xcolor}
\usepackage{enumitem}

\begin{document}
\title{Signatures of Conformal Symmetry in the Dynamics of  Quantum Gases: \\
A Cyclic Quantum State and Entanglement Entropy}

\author{Jeff Maki$^{1)}$}
\author{Fei Zhou$^{2)}$}
\affiliation{1) Pitaevskii BEC Center, CNR-INO and Dipartimento di Fisica, Universit\`a di Trento, I-38123 Trento, Italy \\ 2) Department of Physics and Astronomy, University of British Columbia, 6224 Agricultural Road, Vancouver, BC, V6T 1Z1, Canada}

\begin{abstract}
Conformal symmetry heavily constrains the dynamics of non-relativistic quantum gases tuned to a nearby quantum critical point. One important consequence of this symmetry is that entropy production can be absent in far away from equilibrium dynamics of strongly interacting three-dimensional (3D) and one-dimensional (1D) quantum gases placed inside a soft harmonic trapping potential. This can lead to an oscillatory fully revivable many-body dynamic state, which is reflected in many physical observables. In this article we further investigate the consequences of conformal symmetry on a) the zero-temperature auto-correlation function, 
b) the Wigner distribution function, and c) the Von Neumann entanglement entropy. A direct calculation of these quantities for generic strongly interacting systems is usually extremely difficult.
However, we have derived the general structures of these functions
in the non-equilibrium dynamics when their dynamics are constrained by conformal symmetry. We obtain our results for a) by utilizing an operator-state correspondence which connects the imaginary time evolution of primary operators to different initial states of harmonically trapped gases. While the dynamics of the functions in b) and c) are derived from conformal invariant density matrices.
\end{abstract}

\date{\today}
\maketitle

\section{Introduction}

Far-away from equilibrium quantum dynamics of a generic strongly interacting quantum state has been a highly fascinating topic that is generally an extremely challenging topic to fully explore due to its enormous complexity. The dynamics of strongly interacting quantum states with scale invariant interactions on the other hand can be severely constrained by an emergent dynamic {\em conformal symmetry}. Consequential dynamics due to such a dynamic symmetry have very surprising and distinct features that usually do not appear in generic interacting quantum states. 

The conformal symmetry can naturally appear in non-relativistic scale invariant quantum field theories with Galilean invariance~\cite{Hagen72, Niederer72} and had been applied long ago to understand statistical critical phenomena~\cite{Henkel94}. In the context of quantum gases, the dynamic consequence of scale invariant interaction has also been studied in two spatial dimensions by Pitaevskii and Rosch et.al.~\cite{Rosch97}, although the scale symmetry in two dimensions is only approximate due to quantum anomaly \cite{Ananos03, Olshanii10, Hofmann12}. Later, coordinate re-parametrization and transformations were applied to understand dynamics of unitary gases near Feshbach resonance in a series articles by Castin and Werner~\cite{Castin04, Castin06,Castin12}. At around the same time, Son considered the implications of conformal invariance on hydrodynamics of strongly interacting thermal gases  at high temperatures and the bulk viscosity \cite{Son07}. In addition, the scaling dimensions of primary operators and implications on ground states and excitations in quantum gases were also investigated and discussed in~\cite{Nishida07}. 

Scale and later conformal symmetry have since then been exploited numerous times to examine the different aspects of quantum dynamics of both strongly interacting Bose and Fermi gases in various dimensions \cite{Moroz12, Deng16, Gharashi16,Qu16, Maki18, Deng18, Diao18, Maki19, Dalibard19, Maki20, Maki21,Peng22,Thomas14, Rohringer15,Lv20, Bekassy22, Enss22, Lv22,Maki23, Bekassy23, Wang23}. These studies include an examination of conformal symmetry in the expansion dynamics \cite{Thomas14, Maki18, Maki19}, the dynamics in time-dependent traps \cite{Moroz12, Deng16, Gharashi16, Qu16, Dalibard19, Lv20, Wang23}, the study of breathing dynamics \cite{Castin04, Castin06}, and the existence of large amplitude fully reversible non-equilibrium quantum phenomena or a quantum Boltzmann breather\cite{Maki20,Maki21} that is consistent with the absence of the bulk viscosity from an earlier study \cite{Son07}, short-cuts to adiabaticity and quantum thermodynamics \cite{Deng18, Diao18,Rohringer15, Peng22}, rotating quantum gasses \cite{Bekassy23}, as well as to quantify the role of scale-breaking interactions on the dynamics ~\cite{Maki18,Maki19,Maki20,Maki21, Maki23}.

It is worth remarking that the earlier works on quantum dynamics of three dimensional strongly interacting gases near Feshbach resonance focus on either pure quantum states \cite{Castin04,Castin06,Castin12} or hydrodynamics of strongly interacting thermal gases above the degeneracy temperatures \cite{Son07}. In a series of more recent works, a few new distinct aspects and exciting phenomena in the conformal dynamics that had not been understood before, or had not been connected to dynamic symmetries have been further explored quite intensively and emphasized.

Firstly, the conformal dynamics can occur near any scale invariant fixed points suggested by scale transformations or renormalization group equations (RGEs), independent of particle statistics, high dimensions or low dimensions. Examples are quantum gases near Feshbach resonance, non-interacting quantum gases in 3D and 1D, as well strongly interacting Tonk-Girardeau fixed point in 1D etc \cite{Maki20,Maki21,Maki23}. Notably, we find that far away from equilibrium dynamics near 1D strong coupling limit can be considered to be dual to that near a 3D non-interacting fixed point. Particularly, the scale invariance and consequential conformal symmetry offer a unified picture of quantum dynamics in very different classes of strongly interacting many-body states.

Secondly,  the conformal dynamics occur in strongly interacting systems at both zero temperature, temperatures below the quantum degeneracy temperature, as well as high temperatures where strongly interacting thermal gases emerge, as far as the interactions are scale invariant. In fact, the universal density matrix dynamics are manifestly invariant under conformal transformations at a broad range of temperatures~\cite{Maki19}. Consequently, entropy production dynamics in non-equilibrium dynamics are entirely suppressed resulting in surprising fully reversible far away from equilibrium quantum dynamics in strongly interacting quantum gases in 3D and 1D. Such a quantum phenomenon had been named as a {\em quantum Boltzmann breather} in Ref.~\cite{Maki20} in strongly interacting quantum gases. When taken to high temperatures, the density matrix dynamics indeed also indicate that the bulk viscosity in hydrodynamics of strongly interacting thermal gases has to vanish.

Finally,  the stabilities of these dynamics due to a change in the interaction depends on the nature of the fixed points, i.e.\ the stability depends on both the spatial dimensions and what type of fixed point (strongly interacting vs free particle) the quantum states are nearby \cite{Maki18,Maki19, Maki21}. When stable, conformal symmetry can also appear in the infrared limit as an emergent symmetry in the long wavelength or low frequency limit although the scale symmetry is broken explicitly by strong interactions.

One important consequence of conformal symmetry is the possibility for energy and (thermodynamic) entropy preserving dynamics \cite{Maki19,Maki20}. Such energy and entropy conserving dynamics is so far only known to occur for conformally invariant systems. This feature was illustrated previously in the context of large amplitude oscillations of three-dimensional (3D) Fermi gases with resonantly large s-wave interactions in quenched harmonic oscillator potentials \cite{Maki20}. Provided the initial state is a thermal mixture in a harmonic trap with frequency $\omega_i$, the resulting dynamics in a quenched harmonic trap with frequency $\omega_f < \omega_i$ conserves both energy and entropy and is fully reversible and cyclic. For this situation the gas will expand and contract in the final harmonic trap at a frequency of $2\omega_f$ without end. This motion is the quantum version of the Boltzmann breather \cite{Maki20}.

These large-amplitude oscillations are one signature of the system undergoing a cyclic revival. More remarkably, one can show that the $N$-body density matrix, where $N$ is the number of particles, is completely periodic with frequency $2\omega_f$ \cite{Maki19}. In fact, the time-dependence of the density matrix in position space is a time-dependent rescaling similar to what happens in a simple adiabatic expansion or contraction, but with an important time-dependent gauge factor that represents the coherent global flow and conserves the energy.

In this article we further investigate the consequences of the non-relativistic conformal symmetry and the associated quantum revival in strongly interacting 1D and 3D Fermi gases where the quantum dynamics are cyclic. To this end, we explicitly consider, for the sake of simplicity, the $d$-dimensional strongly interacting Fermi gas placed in a quenched harmonic trap. We then examine how the conformal symmetry and quantum revivals are manifest in the post-quench dynamics. Namely, we investigate:

{\em a) the zero-temperature auto-correlation function for the dynamics of scale invariant quantum gases in quenched harmonic traps (via an imaginary time evolution of a primary operator and the operator-state correspondence);

b) the Wigner distribution function via the density matrix quantum dynamics;

c) the entanglement entropy dynamics.}

The dynamics of these quantities are nigh intractable for generic many-body systems. However, as we will see below, the conformal dynamics and the cyclic nature of the quantum state heavily constrain the dynamics of these quantities to be a simple time-dependent rescaling of their original values.

The remainder of this article is organized as follows. We first review the notion of scale and conformal invariance for unitary Fermi gases in 3D and 1D in Sec.~\ref{sec:review}. We then discuss some general properties of the non-relativistic conformal, or SO(2,1), symmetry in Sec.~\ref{sec:gen_c}. In Sec.~\ref{sec:state_op_corr} we use those general results to  obtain a general state-operator correspondence for harmonically trapped gasses. This correspondence allows us to evaluate the zero-temperature auto-correlation function in Sec.~\ref{sec:auto_corr}. In Sec.~\ref{sec:wigner}, we consider the dynamics of the Wigner distribution function and the momentum distribution. We then examine the Von Neumann entanglement entropy in Sec.~\ref{sec:entropy}. Our final conclusions are presented in Sec.~\ref{sec:conc}.

\section{Hamiltonian for the Strongly Interacting Fermi Gas}
\label{sec:review}

An ideal platform to study conformal symmetry and its implications on dynamics is a spin-$1/2$ Fermi gases in $d=1,3$ spatial dimensions with scale invariant s-wave interactions. In the absence of the trapping potential, the Hamiltonian for the system is:

\begin{align}
H= \int d{\bf r} \ & \psi^{\dagger}({\bf r}) \left( -\frac{1}{2}\nabla^2 \right) \psi({\bf r}) \nonumber \\
&+ \frac{g}{2} \int d{\bf r} \ \psi^{\dagger}({\bf r})\psi^{\dagger}({\bf r}) \psi({\bf r})\psi({\bf r})
\label{eq:H}
\end{align}

\noindent where $\psi^{(\dagger)}({\bf r})$ is the fermionic annihilation (creation) operator, and $g$ is the bare interaction strength. In Eq.~(\ref{eq:H}) we have set $\hbar $ and $m$ to be unity, and have also suppressed the spin indices. 

In 3D the bare coupling constant $g$ is related to the s-wave scattering length according to:

\begin{equation}
\frac{1}{4\pi a} = \frac{1}{g} - \frac{\Lambda}{2\pi^2}
\end{equation}

\noindent where $\Lambda$ is the ultraviolet (UV) cut-off for the theory, and $a$ is the 3D s-wave scattering length. Similarly in 1D the bare coupling constant is related to the 1D s-wave scattering length, which we also define as $a$, via:

\begin{equation}
g= -\frac{2}{a}.
\end{equation}

For finite interaction strengths, the presence of the s-wave scattering length breaks scale symmetry. Hence the dynamics are no longer conformally invariant. Scale invariance reappears in the strongly interacting limit when $a^{-1} = 0$ for 3D and when $a = 0$ for 1D, which can be readily achieved experimentally thanks to Feshbach and confinement induced resonances \cite{Bloch08,Chin10, Olshanii98}.  Unless otherwise stated, we will assume the system resides at the strongly interacting scale invariant point. 

We will be primarily concerned with the dynamics in the presence of a harmonic trapping potential:

\begin{equation}
H_{\omega(t)} =H + \omega^2(t) C
\end{equation}

\noindent where:

\begin{equation}
C = \int d{\bf r} \ \frac{1}{2} r^2 \psi^{\dagger}({\bf r}) \psi({\bf r})
\end{equation}

\noindent and $\omega(t)$ is a general, possibly time-dependent, harmonic trap frequency. 




As a practical set up, We assume that the gas is in thermal equilibrium inside the harmonic trap with frequency $\omega_i$, which can be readily achieved experimentally. At time $t=0$, the system experiences a quench in the trapping frequency to a final value $\omega_f \ll \omega_i$. Provided $\omega_f \ll \omega_i$, the system is in a highly non-equilibrium state and will expand.

\section{General Conformal Symmetry in Strongly Interacting Fermions}
\label{sec:gen_c}

The conformal symmetry discussed in many-body quantum dynamics is closely related to the covariance discussed by Hagen \cite{Hagen72} in the context of a non-relativistic quantum field theory. Consider the case of free space dynamics, $\omega_f = 0$. It was pointed out that the equation of motion for the field $\psi({\bf R},t)$:

\begin{equation}
-i \partial_t \psi({\bf R},t) = \left[H_{\omega_f=0}, \psi({\bf R},t) \right]
\label{eq:psi_eom}
\end{equation}

\noindent as well as the action for a {\em scale invariant} non-relativistic quantum field theory, exhibits a general conformal symmetry, i.e.\ invariance under the following space-time transformation of the coordinates \cite{Son07}:

\begin{align}
{\bf R}  & \rightarrow \tilde{\bf R}=\frac{\bf R}{\lambda(t)}, & dt \rightarrow d\tau= \frac{dt}{\lambda^2(t)}
\label{eq:coord_transf}
\end{align}

\noindent and of the field operator:

\begin{align}
\psi ({\bf R}, t)  & \rightarrow \tilde{\psi}(\tilde{\bf R}, \tau)=\frac{1}{\lambda^{\frac{d}{2}}(t)} \psi(\tilde{\bf R},\tau) \exp [i \frac{{\bf \tilde{R}}^2}{2} \dot{\lambda(t)}^2 ]
\label{eq:psi_transf}
\end{align}

\noindent provided the dynamic re-parametrization factor $\lambda(t)$ is a solution in the following differential equation:

\begin{eqnarray}
\frac{d}{dt} {[\frac{\dot\lambda(t)}{\lambda(t)}]} +[\frac{\dot \lambda(t)}{\lambda(t)}]^2=0.
\label{eq:lambda1}
\end{eqnarray}
Eq.~(\ref{eq:lambda1}) indicates the following solutions:
\begin{align}
\lambda(t)&=1-ct, & -\infty > t > +\infty.
\end{align}
for a general constant $c$ with units of frequency.

\begin{figure}
\includegraphics[scale=0.4]{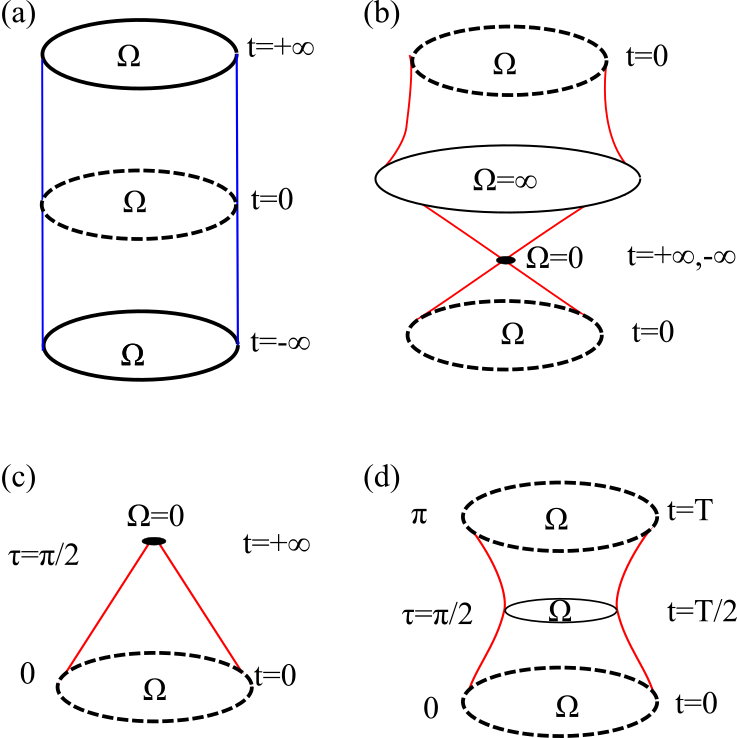}
\caption{
We show the transformation of a finite volume indicated by $\Omega$ along the time domain (defined as the vertical direction). a) represents standard time evolution.
b) is the transformation introduced for the studies of invariance of actions in $(d+1)$ dimension. c) and d) are conformal co-moving coordinates introduced for the studies of $N$-body fermion dynamics in free space c) and in a harmonic trap d).  $\tau(t)$ is the time defined in the conformal co-moving frame (see the main text).}
\label{fig:CFD-shape}
\end{figure}

The space-time mapping in Eqs.~(\ref{eq:coord_transf}-\ref{eq:psi_transf}) describes the dynamics in free space inside a conformal comoving frame also in free space. It is illustrated in Fig.~(\ref{fig:CFD-shape})  b). In this transformation, $t=\pm \infty$ is always mapped into a finite $\tau$ point in the conformal coordinate implying an effective freezing of 
any non-trivial quantum dynamics in the long time limit. 


Often we are interested in the expansion dynamics of quantum gases released from harmonic traps. If we assume the gas is initially in an equilibrium state in a harmonic trap with frequency $\omega_i$,  we can still use the general conformal symmetry to simplify the dynamics. In this case it is much more convenient to force the reparametrization factor, $\lambda(t)$, to be the solution of:

\begin{eqnarray}
\frac{d}{dt} {[\frac{\dot\lambda(t)}{\lambda(t)}]} +[\frac{\dot \lambda(t)}{\lambda(t)}]^2=\frac{\omega_i^2}{\lambda^4(t)}
\label{eq:lambda2}
\end{eqnarray}
The solution to the above equation is:

\begin{align}
\lambda(t)&=\sqrt{1+\omega_i^2 t^2} & \tau(t) &= \frac{1}{\omega_i} \arctan(\omega_i t).
\label{eq:coord_transf_2}
\end{align}

\noindent In the conformal comoving frame defined by Eq.~(\ref{eq:coord_transf_2}), the free-space dynamics are mapped onto the dynamics inside in a harmonic trapping potential with frequency $\omega_i$ \cite{Maki19}. This emergent harmonic trapping potential can be thought of as a fictitious force in the non-inertial conformal comoving frame. In this frame, the long-time dynamics are simple as: $\lambda(t) \approx \omega_i t$ and $\tau(t) \approx \pi/2 +O(1/(\omega_i t)$. These conformal coordinates effectively compactify quantum dynamics in the time domain (see Fig.~(\ref{fig:CFD-shape}) c.)

The general conformal transformation is also useful in describing the dynamics in a harmonic trap with frequency $\omega_f$. Again, we will consider the case where the gas is initially in equilibrium in a harmonic trap with frequency $\omega_i$. In this case the time-dependent scaling factor satisfies:

\begin{eqnarray}
\frac{d}{dt} {[\frac{\dot\lambda(t)}{\lambda(t)}]} +[\frac{\dot \lambda(t)}{\lambda(t)}]^2 +\omega^2_f=\frac{\omega_i^2}{\lambda^4(t)}.
\label{eq:lambda2}
\end{eqnarray}

\noindent The solution to Eq.~(\ref{eq:lambda2}) has the following form:

\begin{eqnarray}
\lambda(t) &=&\sqrt{\cos^2 (\omega_f t) +\frac{\omega^2_i}{\omega^2_f} \sin^2(\omega_f t)} \nonumber \\
\tau(t) &=&\frac{1}{\omega_i}\arctan\left(\frac{\omega_i}{\omega_f}\tan( \omega_f t)\right) 
\label{eq:lambda_quench}
\end{eqnarray}

\noindent This coordinate transformation maps the dynamics in a harmonic trap with frequency $\omega_f$ into a comoving frame where the system is inside the original harmonic trap of frequency $\omega_i$. In this conformal comoving frame, $\tau(t)$ is effectively compactified on the domain  $[0,\pi]$ with $t=+\pi/2\omega_f$ being mapped into $\tau=\pi/2$. This situation is illustrated in Fig.~(\ref{fig:CFD-shape}) d).

In general this general conformal transformation can be applied to the dynamics of arbitrary time-dependent harmonic trapping potentials. In this case the relevant equation is the Ermakov equation:
\begin{eqnarray}
\frac{d}{dt} {[\frac{\dot\lambda(t)}{\lambda(t)}]} +[\frac{\dot \lambda(t)}{\lambda(t)}]^2 +\omega^2(t) = \frac{\omega_i^2}{\lambda^4(t)}.
\label{eq:lambda3}
\end{eqnarray}
In this situation, the harmonic trap in the comoving frame has a frequency $\omega_i$. This has been used previously to study scale invariant quantum gases in time-dependent harmonic traps \cite{Moroz12, Deng16, Gharashi16, Qu16, Dalibard19, Lv20}. 

We note that the dynamics of the scaling factor is identical to a hydrodynamic analysis of the expansion dynamics. For completeness we  present this analysis in Appendix \ref{appendix:hydro_amp}. Below we will exploit the compactified effective time coordinate, $\tau$, to understand the dynamics of strongly interacting scale invariant Fermi gases.


\section{State-Operator Correspondence for Quenched Harmonic Traps}
\label{sec:state_op_corr}

Another important consequence of conformal symmetry is related to the spectrum of the Hamiltonian, $H_{\omega}$. In particular, one can show that the energy spectrum of the Hamiltonian $H_{\omega}$ consists of a set of conformal towers. The eigenstates within a conformal tower are evenly spaced by an amount $2 \omega$ \cite{Castin06,Nishida07}. However, the difference in the ground state energies of two conformal towers are not generally commensurate in the strongly interacting limit \cite{Nishida07}. 

The ground states of each conformal tower is special as it can be constructed from a primary operator, $O_p$ which satisfy: $[C,O_p] = [K_i, O_p]=0$, where $C$ is the generator of conformal transformations, and $K_i$ is the generator of Galilean boosts along the $i= x,y,z$ direction. For our discussions on more general nonequilibrium dynamics in a soft trap with frequency $\omega_f$, it is imperative that we are able to connect $O_p$ to a wide range of initial states that can be easily prepared experimentally.
Here we show that the ground state  of a trapped gas with frequency $\omega_i$ can be obtained through the imaginary time evolution of the primary operator in a harmonic trap with frequency $\omega_f$ via the state-operator correspondence.


Let $|\psi_0\rangle$ be an initial state prepared as the ground state of the Hamiltonian $H_{\omega_i}$ in experiments. We will show that the the ground state of the $H_{\omega_i}$ is related to imaginary time-evolution in the trap $H_{\omega_f}$:

\begin{align}
|\psi_0\rangle &= e^{-H_{\omega_f} \tau} O_p^{\dagger} | vac\rangle,
 &\tanh (\omega_f \tau) &=\frac{\omega_f}{\omega_i}
\label{eq:psi0}
\end{align} 
\noindent where $\tau$ is the amount of imaginary time needed to achieve a desired initial state $|\psi_0\rangle$ from $O_p^{\dagger}$ acting on the many-body vacuum, $|vac\rangle$. 
Note $\tau$ is a function of $\omega_f$ and $\omega_i$ given by the solution to the second equation above.

To prove Eq.~(\ref{eq:psi0}) it is sufficient to show that $|\psi_0\rangle$ is indeed an eigenstate of $H_{\omega_i}$ for a given $\tau(\omega_i,\omega_f)$. Further conformal symmetry arguments require this state to be the ground state. 
Consider the following equation that is identical to Eq.~(\ref{eq:psi0}):

\begin{equation}
H_{\omega_i} |\psi_0 \rangle = e^{-H_{\omega_f}\tau} e^{H_{\omega_f}\tau} H_{\omega_i} e^{-H_{\omega_f}\tau} O^{\dagger}|vac\rangle
\label{eq:psi1}
\end{equation}

\noindent Eq.~(\ref{eq:psi1}) can be evaluated analytically using the SO(2,1) algebra in Table \ref{tab:commutation_relations}. From the SO(2,1) algebra one can show that:

\begin{align}
e^{H_{\omega_f}\tau} H_{\omega_i} e^{-H_{\omega_f}\tau} &= \left(1-\frac{\omega_i^2-\omega_f^2}{\omega_f^2}\sinh^2(\omega_f \tau)\right)  H_{\omega_f} \nonumber \\
&-i\frac{\omega_i^2-\omega_f^2}{2\omega_f} \sinh(2\omega_f \tau) D \nonumber \\
&+ (\omega_i^2-\omega_f^2) \cosh(2 \omega_f \tau) C
\end{align}

\noindent To simplify things further we use the properties of the primary operator, namely its commutation relations: $[D, O_p^{(\dagger)}] = i \Delta_{O_p} O_p^{(\dagger)}$ where $\Delta_{O_p}$ is the scaling dimension of $O_p^{(\dagger)}$ and $[C,O_p^{(\dagger)}]=0$:

\begin{align}
H_{\omega_i} |\psi_0 \rangle &= \left(1-\frac{\omega_i^2-\omega_f^2}{\omega_f^2}\sinh^2(\omega_f \tau)\right)  H_{\omega_f} |\psi_0\rangle \nonumber \\
&+\frac{\omega_i^2-\omega_f^2}{2\omega_f} \sinh(2\omega_f \tau) \Delta_{O_p} |\psi_0\rangle
\label{eq:op_state_2}
\end{align}

In order to remove the spurious term related to $H_{\omega_f}$, we set:

\begin{align}
\sinh(\omega_f \tau) &= \frac{\omega_f}{(\omega_i^2-\omega_f^2)^{\frac{1}{2}}}, & & \mbox{or} & \tanh (\omega_f \tau)& =\frac{\omega_f}{\omega_i}.
\label{eq:tau_def}
\end{align}

\noindent In this way Eq.~(\ref{eq:op_state_2}) simplifies to: 

\begin{equation}
H_{\omega_i} |\psi_0 \rangle = \Delta_O \omega_i |\psi_0\rangle
\end{equation}

\noindent which by construction is the ground state of $H_{\omega_i}$.

Previously this state-operator correspondence was shown to relate the ground state of a trapped gas to imaginary time evolution in free space with $H_{\omega_f=0}$ \cite{Nishida07}.
We note that in the limit $\omega_f \to 0$, i.e. for the case of the final Hamiltonian being in free-space, our result states $\tau = 1/\omega_i$, which is the result for free space evolution.

\begin{figure}
\includegraphics[scale=0.45]{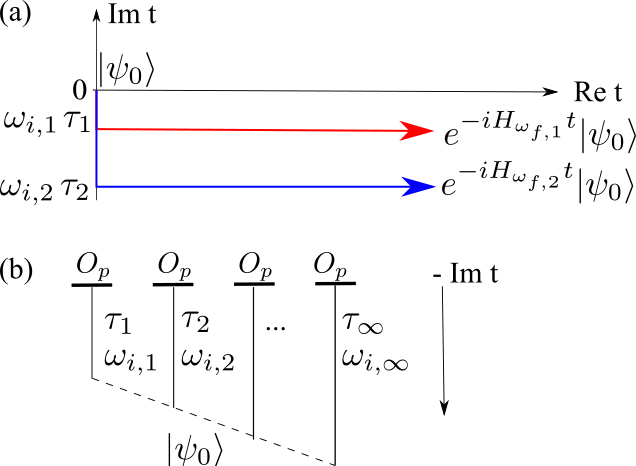}
\caption{Non-equilibrium quantum dynamics via imaginary time evolution of a primary conformal operator. In a), different paths in the plane of $Re t -Im t$ can be used to represent various quantum dynamics under the actions of {\em different} Hamiltonians but with the same initial state.
They can also represent unitary evolution under the action of the same Hamiltonian but {\em different} initial states. In b), we schematically show that imaginary time evolution of the same operator $O_p$ under the action of {\em different} Hamiltonians can result in the same state $|\psi_0\rangle$ if a proper imaginary time duration is chosen for each distinct Hamiltonian. See the main text in Section \ref{sec:state_op_corr} for details.}
\label{fig:CDF-Itime}
\end{figure}

As in the case of free space, the dynamics of the state  $|\psi_0 \rangle$ following the Hamiltonian $H_{\omega_f}$ is a simple time-dependent rescaling of the many-body wavefunction. This can be readily seen by examining:

\begin{equation}
|\psi_0\rangle(t) = e^{-i H_{\omega_f} t} |\psi_0\rangle = e^{-i H_{\omega_f} (t-i \tau)} O^{\dagger}_p |vac \rangle
\end{equation}

\noindent In other words, the dynamics of a primary state is given by a time evolution of $O_p^{\dagger}|vac\rangle$ over complex time $t-i\tau$. 

By changing the amount of imaginary time evolution, i.e. $\tau$, one can access different final Hamiltonians via Eq.~(\ref{eq:tau_def}). Similarly to obtain different times, one tunes the real part of the time evolution. This is shown schematically in Fig.~(\ref{fig:CDF-Itime}) a). The converse is also true, by fixing the final Hamiltonian, and by tuning the imaginary time evolution, one can have access to different initial states. Thus by suitably adjusting the imaginary time-evolution, one can examine a wide range of initial conditions and dynamics.

Following the same argumentation, one finds that at each moment in time $|\psi_0\rangle(t)$ is an exact eigenstate of a Hamiltonian of the form:

\begin{equation}
H_{\omega_i}\left(\frac{\bf r}{\lambda(t)}, \lambda(t) {\bf p} - \dot{\lambda}(t)\right) |\psi_0 \rangle(t) = \Delta_{O_p} \omega_i |\psi_0\rangle(t) 
\end{equation}

\noindent where $\lambda(t)$ is given by Eq.~(\ref{eq:lambda_quench}), and we have highlighted the position and momentum dependence of the Hamiltonian as: $H_{\omega_i}({\bf r}, {\bf p})$. This is equivalent to the statement of scaling dynamics proposed in the context of the general conformal symmetry of the equation of motion. This result was previously derived for the case of free-space dynamics \cite{Maki20}, but equally applies to any initial and final trap provided $\tau$ is chosen accordingly.

\section{Zero-Temperature Auto-Correlation Function in Quenched Harmonic Traps}
\label{sec:auto_corr}

From the state-operator correspondence, we can present a different illustration of the conformal dynamics. Consider the zero-temperature generalized auto-correlation function in a quenched harmonic trap:

\begin{equation}
{G}({\bf R},t) = \left\langle \psi_0 \right| e^{-iH_{\omega_f}t} e^{i {\bf P \cdot R}} \left|\psi_0 \right\rangle 
\label{eq:def_auto_corr}
\end{equation}

\noindent where $|\psi_0\rangle$ is the ground state of the Hamiltonian $H_{\omega_i}$. For a general discussion, we also define the momentum operator in the ith direction as $P_i$ so that we can also translate
the center of $|\psi_0 \rangle$ to point ${\bf R}$ in space.

The auto-correlation function defined in Eq.~(\ref{eq:def_auto_corr}) is a special case of the Loschmidt echo evaluated at zero-temperature. In general the Loschmidt echo quantifies the overlap between a given initial state time evolved according to two different Hamiltonians, $H_1$ and $H_2$:

\begin{equation}
\mathcal{L}(t) = \langle \psi_0 | e^{i H_2 t} e^{-i H_1 t} | \psi_0 \rangle.
\label{eq:Loschmidt}
\end{equation}

\noindent Eq.~(\ref{eq:Loschmidt}) is exactly  Eq.~(\ref{eq:def_auto_corr}), with $H_1 = H_{\omega_f}$ and $H_2 = H_{\omega_i}$, except we further separate these two states in position space by ${\bf R}$. 

First, we can write the auto-correlation function in terms of a correlation function of the primary operator:

\begin{equation}
G({\bf R},t) = \langle vac | O_p({\bf R}, t-i \tau) O_p^{\dagger}(0,i \tau) |vac \rangle
\end{equation}

\noindent where we have defined:

\begin{equation}
O_p^{(\dagger)}({\bf R},t) = e^{-i{\bf P \cdot R}} e^{i H_{\omega_f}t} O_p^{(\dagger)} e^{-i H_{\omega_f} t} e^{i {\bf P \cdot R}} 
\end{equation}

 The non-relativistic conformal symmetry completely restricts the form of such correlation funcitons \cite{Nishida07,Golkar14, Goldberger15}. Consider:

\begin{equation}
0 = \langle vac | [C, O_p({\bf R},t-i \tau) O_p^{\dagger}(0,i \tau)] | vac \rangle
\label{eq:conformal_constraint}
\end{equation}

\noindent since $C|vac\rangle = 0$. As discussed in Appendix \ref{app:auto_corr}, it is possible to evaluate Eq.~(\ref{eq:conformal_constraint}) from the Schroedinger algebra in Tab.~\ref{tab:commutation_relations}:

\begin{align}
0= & \left[\frac{\sin^2(\omega_f (t-i\tau))}{\omega_f^2} \partial_{t} + \frac{\sin(2\omega_f (t-i\tau))}{2\omega_f }\left({\bf R \cdot \nabla_{\bf R}} + \Delta_{O_p}\right) \right. \nonumber \\
& \left. + i N_{O_p} \frac{R^2}{2} \cos(2\omega_f (t-i\tau)) \right]\
G({\bf R},t)
\label{eq:diff_eqn_auto_correlation}
\end{align}

\noindent where $\Delta_O$ and $N_O$ are the scaling dimension and number of the primary operator $O_p^{\dagger}$: $[D,O_p^{\dagger}] = i \Delta_{O_p} O_p^{\dagger}$ and $[N,O_p^{\dagger}] = N_{O_p} O_p^{\dagger}$.
Similarly, $[{\bf K}, O_p]=0$ leads to the following dynamic equation,

\begin{align}
0= \left[ \frac{\sin(\omega_f (t-i\tau))}{\omega_f } {\nabla}_{\bf R}
 + i N_{O_p} {\bf R} \cos(\omega_f (t-i\tau)) \right] G({\bf R},t).
\label{eq:diff_eqn_auto_correlation1}
\end{align}

The solution to Eqs.~(\ref{eq:diff_eqn_auto_correlation}-\ref{eq:diff_eqn_auto_correlation1}) is unique and gives us the final form of the zero-temperature auto-correlation function:

\begin{align}
G({\bf R},t) &= \left(\frac{\omega_f^2}{\sin^2\left(\omega_f (t-i\tau)\right)}\right)^{\frac{\Delta_{O_p}}{2}} \nonumber \\
&\times\exp\left[-i \frac{N_{O_p}}{2} R^2 \frac{\omega_f}{\tan\left(\omega_f (t-i\tau) \right)}\right]
\label{eq:auto_corr_sol}
\end{align}

Eq.~(\ref{eq:auto_corr_sol}) is periodic with a frequency of $2\omega_f$. This is a signature of both the lack of entropy production in the conformal dynamics, and the evenly spaced spectrum of the conformal towers. This result is  also consistent with the $N$-body density-matrix analysis which states that the time-evolved many-body wavefunction returns to itself at time $t = \pi/\omega_f$. We note that the explicit value of $G({0,0})$, is related to the normalization of the many-body wavefunction. 

Eq.~(\ref{eq:auto_corr_sol}) is also valid in the limit of free-space expansion, $\omega_f \to 0$ and $\omega_f t \ll \pi$. In this case, the auto-correlation function decays as a power-law in time, consistent with Galilean invariance:

\begin{align}
\lim_{\omega_f \to 0} G\left({\bf R}, t\ll \frac{\pi}{\omega_f}\right) \approx \frac{1}{t^{\Delta_{O_p}}} e^{-i \frac{N_{O_p}}{2}\frac{R^2}{t}}
\end{align}

\noindent where we have disregarded the initial conditions for comparison to previous results \cite{Nishida07, Golkar14,Goldberger15}.


\begin{table}[]
\begin{tabular}{|c|c|c|c|c|c|}
\hline
$\left[A, B\right]$ & $H$ & $C$ & $D$ & $P_i$ & $K_i$ \\ \hline
$H$ & $0$ & $-iD$ & $-2iH$  & $0$ & $-i P_i$\\ \hline
$C$ &  $i D$ & $0$ & $2iC$ & $i K_i$ & $0$ \\ \hline
$D$ & $2iH$ & $-2iC$ & $0$ & $i P_i$& $-i K_i$\\ \hline
$P_j$ & $0$ & $-i K_j$ &  $-i P_j$ & $0$ & $-i \delta_{i,j}N$\\ \hline
$K_j$ &  $i P_j$ & $0$ & $i K_j$ & $i \delta_{i,j} N$ & $0$ \\ \hline
\end{tabular}
\caption{Commutation relations for the Schroedinger algebra. The rows and columns are the operators $A$ and $B$ respectively.  The operators shown above are derived in the main text. The Schroedinger algebra also contains the angular momentum operator, which we ignore as we focus on isotropic systems. We also define $N$ as the number operator, which commutes with all the above operators. The SO(2,1) algebra repersents the sub-algebra spanned by $H$, $C$, and $D$. } 
\label{tab:commutation_relations}
\end{table}

\section{Wigner Distribution}
\label{sec:wigner}

A more experimentally practical signature of the cyclic quantum dynamics is to examine the Wigner distribution function, and in a related manner, the momentum distribution. The Wigner distribution function is related to the one-body density matrix:

\begin{equation}
f({\bf R},{\bf k};t) =\int d^d{\bf r} e^{-i {\bf k}\cdot{\bf r}} \rho_1\left(\frac{{\bf R}}{2} +{\bf r}, \frac{{\bf R}}{2}-{\bf r};t\right)
\label{eq:wigner}
\end{equation}  
where $\rho_1\left(\frac{{\bf R}}{2} +{\bf r}, \frac{{\bf R}}{2}-{\bf r};t\right)$ is the one-body density matrix, ${\bf R}=({\bf r}+{\bf r}')/2$ is the center of mass coordinate and  ${\bf k}$ is the relative momentum. 

Previously, we proved that the $N$-body density matrix for the system can be written in real space as \cite{Maki20}:

\begin{align}
\rho_N(\lbrace {\bf r}_i\rbrace,\lbrace {\bf r}'_j\rbrace, t) &= \langle \psi_0 | \prod_{j=1}^N \psi^{\dagger}({\bf r}_i', t) \prod_{i=1}^N \psi({\bf r}_j, t) |\psi_0 \rangle \nonumber \\
&= \frac{1}{\lambda^{dN}(t)} \exp\left[i \frac{\dot{\lambda}(t)}{\lambda(t)} \sum_{i=1}^N\frac{r_i^2-r_i'^2}{2}\right] \nonumber \\
& \rho_N\left(\left\lbrace \frac{{\bf r}_i}{\lambda(t)} \right\rbrace,\left\lbrace \frac{ {\bf r}'_i}{\lambda(t)} \right\rbrace, 0\right)
\label{eq:dm_final}
\end{align}

\noindent with $\lbrace {\bf r_i} \rbrace$ is the set coordinates for $N$ particles. Eq.~\eqref{eq:dm_final} is exact if the initial density matrix is the one of an equilibrium state in a harmonic trap. For more general initial states, Eq.~\eqref{eq:dm_final} represents asymptotic dynamics of a strongly interacting quantum gas when $t \rightarrow \infty$.

This density matrix result naturally extends to the one-body density matrix. Substituting Eq.~(\ref{eq:dm_final}) into Eq.~(\ref{eq:wigner}) yields a surprisingly simple result:

\begin{equation}
f({\bf R}, {\bf k};t) = f\left(\frac{{\bf R}}{\lambda(t)}, {\bf k} \lambda(t) - {\dot \lambda(t)} {\bf R}; 0\right)
\label{eq:Wigner_time}
\end{equation} 

\begin{figure}
\includegraphics[scale=0.55]{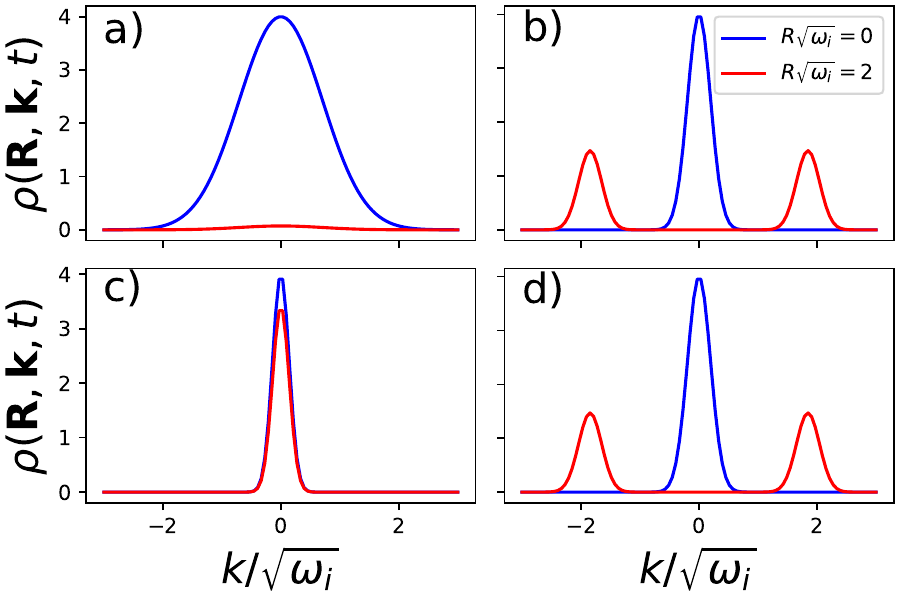}
\caption{Schematic for the Wigner distribution, Eq.~(\ref{eq:Wigner_time}), as a function of momentum $k$ for two center-of-mass coordinates, $R$ (blue: $R=0$, and red: $R = 2/\sqrt{\omega_i}$). The plots $a)-d)$ correspond to times $ t/T_f = 0, 1/4, 1/2, 3/4$, respectively. When $R=0$ the dynamics are a simple rescaling, while for finite $R$ the dynamics also  involves a time-dependent shift of the center of the Wigner distribution.} 
\label{fig:wigner}
\end{figure}

The dynamics of the Wigner distribution function, Eq.~(\ref{eq:Wigner_time}) is completely constrained by the conformal symmetry. The dynamics of the Wigner distribution function are shown in Fig.~(\ref{fig:wigner}). The dynamics contain a time-dependent rescaling of the position and momentum coordinates alongside a translation in momentum space by $\dot{\lambda}(t) {\bf R}$. Such dynamics are again completely periodic at a frequency $2\omega_f$ and is another manifestation of the {\em Quantum Boltzmann breather} discussed in Ref.~\cite{Maki20}.

One interesting observation is that Eq.~\eqref{eq:Wigner_time} appears to be a solution to the following transport equation:
\begin{equation}
0=\left[\partial_t + {\bf k} \cdot \nabla_{\bf R} -\omega_f^2 {\bf R} \cdot \nabla_{\bf k}\right] f({\bf R}, {\bf k}, t)
\label{eq:boltzmann}
\end{equation}
if at $t=0$ the Wigner function is of an equilibrium form.

Eq.~\eqref{eq:boltzmann} resembles the quantum kinetic equation, in the Fermi liquid theory, for the quasiparticle distribution function in the collisionless limit \cite{Pitaevskii12, Baym}. This is quite surprising as our examination of the Wigner distribution function contains the full quantum dynamics in the strongly interacting regime {\em and is valid at both below and above quantum degeneracy temperatures}. In fact, the temperature during the course of dynamics is simply related to the initial one via a proper time-dependent rescaling.
   
A quantum kinetic description for a generic strongly interacting system is expected to be much more complicated than Eq.~\eqref{eq:boltzmann}. In particular, one expects there to be a finite collision integral strongly depending on temperatures, as well as renormalization effects of the quasiparticle velocity due to self-energy effects etc. These features appear to be entirely absent in Eq.~\eqref{eq:boltzmann}. Similarly, Eq.~\eqref{eq:boltzmann} is almost identical to the set of GHD equations for the rapidity distribution with hardcore contact interaction in 1D \cite{Castro16, Bertini16, Bulchandani17, Caux19, Bouchoule22, Essler22}. 

Also Eq.~\eqref{eq:boltzmann} resembles the classical kinetic theory for the phase space distribution function. The classical kinetic theory can also support a undamped motion known as the Boltzmann breather in {\em a weakly interacting gas} \cite{Boltzmann}. This motion was later observed in thermal Bose gases in Ref.~\cite{Lobser15}. In the Boltzmann breather, the classical distribution function maintains its equilibrium form and undergoes a rescaling motion that is similar to what is suggested by Eq.~\eqref{eq:Wigner_time}. Notably the collision integral vanishes identically, and the motion of the distribution function follows Eq.~\eqref{eq:boltzmann}. From the point of view of conformal dynamics, both the non-interacting and strongly interacting scale invariant fixed points shall naturally exhibit conformal symmetry which supports the undamped motion.

A natural corollary of Eq.~(\ref{eq:Wigner_time}) is that the momentum distribution:

\begin{equation}
n(k,t) = \int d^d{\bf R} f({\bf R}, {\bf k},t)
\end{equation}
is also completely oscillatory in conformal dynamics at a frequency of $2\omega_f$. These oscillations in the momentum distribution were originally posited in Ref. \cite{Minguzzi05} and seen experimentally \cite{Wilson20} for 1D Bose gases in the Tonks-Girardeau limit. Here we state that this phenomenon is more generally associated with the conformal dynamics of scale invariant quantum gases expanding in harmonic traps, and thus can be observed in higher-dimensions, for example the 3D spin-$\frac{1}{2}$ Fermi gases at a Feshbach resonance. And it shall occur at a wide range of temperatures both below and above degeneracy temperatures.

\section{Von Neumann Entropy and conformal dynamics}
\label{sec:entropy}

The fact that conformal dynamics are completely periodic at a frequency $2\omega_f$ means that there is no thermodynamic entropy production. This is evident from the scaling dynamics of the density matrix in Eq.~(\ref{eq:dm_final}). However, one can ask about the dynamics of the entanglement entropy. Can the reversible nature of conformal dynamics be detected in the dynamics of the entanglement entropy? Indeed, as we will now show, conformal symmetry also constrains the entanglement entropy.

The starting point is the $N$-body density matrix defined in Eq.~(\ref{eq:dm_final}).
To calculate the entanglement entropy we will partition real space into two subspaces, $\mathcal{A}$ and $\mathcal{B}$. Since particles, can either be in either $\mathcal{A}$ or $\mathcal{B}$, it is possible to write:
\begin{align}
\rho_N(\lbrace {\bf r}_i\rbrace,\lbrace {\bf r}'_j\rbrace, t) = &\prod_i \left(\theta(\vec{r}_i\in \mathcal{A}) + \theta(\vec{r}_i \in \mathcal{B})\right)\nonumber \\
&\prod_j \left(\theta(\vec{r}'_j\in \mathcal{A}) + \theta(\vec{r}'_j \in \mathcal{B})\right) \nonumber \\
&\rho_N(\lbrace {\bf r}_i\rbrace,\lbrace {\bf r}'_j\rbrace, t) 
\end{align}
where $\theta(\vec{r}_i\in \mathcal{A}(\mathcal{B}))$ is unity if $\vec{r}$ is in $\mathcal{A}$ ($\mathcal{B}$) and is zero otherwise. We focus on terms that have fixed particle number in each definite subspace. Then by performing the trace over $\mathcal{B}$, we obtain the reduced density matrix for $\mathcal{A}$, see Fig~(\ref{fig:entanglement}). 

The density matrix in subspace $\mathcal{A}$ can be written as a direct sum of density matrices describing the state of the system with $k$ particles in the subspace \cite{Sugishita21}:

\begin{figure}
\includegraphics[scale=0.5]{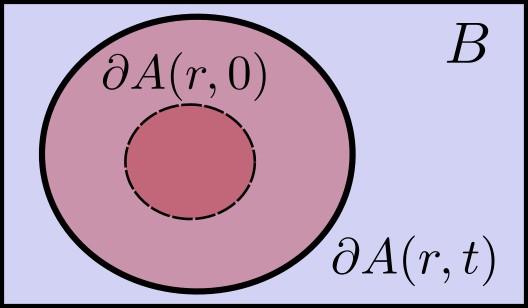}
\caption{Schematic for the entanglement Entropy. The region $A$ is shown by the red region, while $B$ is the blue region. The boundary of $A$ is denoted by $\partial A(r,t)$. The dynamics of the entanglement entropy are entirely coded in the motion of the boundary: $\partial A(r,t) = \partial A(r/\lambda(t),0)$.}
\label{fig:entanglement}
\end{figure}

\begin{equation}
\rho_{\mathcal{A}} = \bigoplus_{k=0}^N p_k \rho_{\mathcal{A},k}
\label{eq:decomp}
\end{equation}

\noindent where $p_k$ is the probability of finding $k$ atoms in $\mathcal{A}$:

\begin{align}
p_{k} = {N \choose k} \int\lbrace & d{\bf r}_{\mathcal{A},i}\rbrace\int\lbrace d{\bf r}_{\mathcal{B},j} \rbrace \nonumber \\
&\rho_N(\lbrace {\bf r}_{\mathcal{A},i} , {\bf r}_{\mathcal{B},j}\rbrace,\lbrace {\bf r}_{\mathcal{A},i}, {\bf r}_{\mathcal{B},j}\rbrace,t) 
\end{align}

\noindent and $\rho_{\mathcal{A},k}$ is the normalized density matrix of subspace $\mathcal{A}$ with $k$ atoms:

\begin{equation}
\rho_{\mathcal{A},k} = \frac{1}{p_k}\int\lbrace d{\bf r}_{\mathcal{B},j} \rbrace\rho_N(\lbrace {\bf r}_{\mathcal{A},i} , {\bf r}_{\mathcal{B},i}\rbrace,\lbrace {\bf r}_{\mathcal{A},j}', {\bf r}_{\mathcal{B},j}\rbrace,t) 
\end{equation}

\noindent In the above formulae ${\bf r}_{\mathcal{A},i}$ is the position of the $i$th particle in $\mathcal{A}$ ($i = 1,2,...k$), and similarly ${\bf r}_{\mathcal{B},j}$ is the position of the $j$th particle in $\mathcal{B}$ ($j=1,2,...N-k$). 

The probability $p_k$ has two main components. The first is a  factor of ${n \choose k}$ which represents the combinatorics of having $k$ particles out of $N$ in $\mathcal{A}$. The second part is the trace of the $N$-body density matrix provided $k$-particles are in $\mathcal{A}$. For translational invariant systems, this is intuitively related to the probability of finding $k$ particles in $\mathcal{A}$ and the rest in $\mathcal{B}$. The probability of finding a single particle in either $\mathcal{A}$ or $\mathcal{B}$ is $\approx V_{\mathcal{A}(B)}/V$ where $V_{\mathcal{A} (B)}$ is the volume of  $\mathcal{A}$ or $\mathcal{B}$ and $V= V_{\mathcal{A}} + V_{\mathcal{B}}$ is the total volume. This implies the following expected behaviour of the probability:
\begin{align}
p_k &\approx {N \choose k} \frac{V_{\mathcal{A}}^k V_{\mathcal{B}}^{N-k}}{V^N}  & \sum_k p_k = 1.
\end{align} 
Although this was derived in free space, one can prove that $\sum_k p_k = 1$ holds more generally \cite{Sugishita21}.

Given the decomposition of $\rho_{\mathcal{A}}$ in Eq.~(\ref{eq:decomp}), the von Neumann, or entanglement, entropy for $\mathcal{A}$:

\begin{equation}
S_{vN}(t) = - Tr_{\mathcal{A}} \left[ \rho_{\mathcal{A}} \log \rho_{\mathcal{A}}\right]
\label{app:vn_ent}
\end{equation}

\noindent simplifies into a classical and a quantum piece:

\begin{align}
S(t) &= S_c + S_q(t), \nonumber \\
\label{eq:classical_ent}
S_c  &= -\sum_k p_k \ln p_k, \\
\label{eq:quantum_ent}
S_q(t) &= -\sum_k p_k Tr_{\mathcal{A},k} \left[\rho_{\mathcal{A},k} \cdot \log\rho_{\mathcal{A},k}\right].
\end{align}

\noindent The classical piece is just the Shannon entropy for finding $k$ particles in $\mathcal{A}$ with probability $p_k$, while the quantum piece gives the entanglement entropy of the $k$ particles in $\mathcal{A}$.

The dynamics of the entanglement entropy is contained in $S_q(t)$, which depends on the reduced density matrices, $\rho_{\mathcal{A},k}$. The dynamics of these reduced density matrices immediately follows from Eq.~(\ref{eq:dm_final}); they are equivalent to a time-dependent rescaling of the position coordinate up to a gauge transformation. Thus one expects the total entanglement entropy to be conserved, or at least oscillatory. This depends on the choice of the subspace $\mathcal{A}$. 

Let the boundary of the $\mathcal{A}$ be denoted as $\partial \mathcal{A}({\bf r})$. For general $\partial \mathcal{A}({\bf r})$, the entanglement entropy will oscillate at a frequency of $2\omega_f$, as the dynamics of the density matrices are oscillatory at this frequency. If one considers the comoving frame moving with the scale factor $\lambda(t)$, the boundary of $\mathcal{A}$ becomes time dependent: $\partial \mathcal{A}({\bf r}/\lambda(t))$. In the comoving frame, the oscillations of the entanglement entropy are due to the oscillations of the boundary, see Fig.~(\ref{fig:entanglement}). Naturally if one judicially chooses $\mathcal{A}$ to be a semi-infinite plane with its boundary at the origin, the entanglement entropy will be conserved at all times as $\partial \mathcal{A}({\bf 0}/\lambda(t)) = \partial \mathcal{A}({\bf 0})$. In both cases such trivial dynamics of the entanglement entropy are related to conformal symmetry and the motion of the boundary of $\mathcal{A}$, not necessarily related to information scrambling of the whole system.

\section{Conclusion}
\label{sec:conc}
In this article we have examined several signatures of conformal symmetry on dynamics. First we examined the operator state correspondence for harmonically trapped gases, which connects the imaginary time evolution to different initial states and real time evolution to different dynamical states. We employed this result to examine the oscillatory motion of a) the zero-temperature auto-correlation function, b) the Wigner distribution function, and c) the entanglement entropy. In all these quantities, the conformal symmetry completely restricts the dynamics, making them related to a simple time-dependent rescaling of their initial values. This is a huge simplification of the original problem which would require a microscopic calculation of the dynamics. 

This revival is not related to the many-body Poincare recurrence time which is exponentially long and not a perfect revival \cite{Bocchieri57}, nor is it related to the one-body recurrence as the one-body spectrum is evenly spaced. Rather this many-body revival is indeed exact provided the system is initially in thermal equilibrium, and is a consequence of the non-relativistic conformal symmetry and the dynamics of the resulting conformal tower states \cite{Castin04, Castin06, Nishida07}. For more generic initial states, our results represent long time asymptotics of general expansion dynamics. This has been emphasized in Ref.~\cite{Maki19}.

Since this simplification of the dynamics depends on the symmetry alone, it is quite general and applies to arbitrary scale invariant quantum gas. Thus these results not only apply to the strongly interacting gas in 1D and 3D, but also the non-interacting, regardless of dimensionality. As discussed in a recent work by the authors \cite{Maki23}, these results also apply to the case of expansion dynamics in the presence of irrelevant perturbations, like finite range effects in the two-body scattering, three-body interactions, and in 1D, small deviations from strong scale invariant interactions. In these systems, the effects of breaking scale invariance occur in a limited short time window when the gas is most dense, and as such take many cycles to accumulate.  For these systems conformal symmetry can still be readily applied and provides a clear path towards an analytical understanding of far-away-from-equilibrium dynamics.

The authors would like to thank Randy Hulet and Shizhong Zhang for useful discussions. This project was partially supported by the NSERC (Canada) Discovery Grant.

\appendix

\numberwithin{equation}{section}
\renewcommand\theequation{\Alph{section}.\arabic{equation}}

\section{Conformal Dynamics in a Two-Quench Protocol}
\label{appendix:hydro_amp}

In this section we review the dynamics of scale invariant and nearly scale invariant quantum gasses inside a harmonic trap with time-dependent frequency:

\begin{equation}
\omega(t) = 
\begin{cases}
\omega_i & t<0 \\ 
\omega_f & 0< t<t_h \\
\omega_i & t>t_h
\end{cases}
\label{eq:omega(t)}
\end{equation}
where $t_h$ is the hold time. Initially we assume the gas is in thermal equilibrium. After the first quench, $0<t<t_h$, the gas undergoes conformal dynamics which can be described by the scale factor in Eq.~(\ref{eq:lambda_quench}). Equivalently, one can use the hydrodynamic equations of motion \cite{Thomas14, Maki20} to obtain the following differential equation for the moment of inertia:

\begin{align}
\frac{d^2}{dt^2} \langle C \rangle(t) &=2\left(\omega_i^2 + \omega_f^2 \right) \langle C \rangle(0) - 4 \omega_f^2 \langle C \rangle(t)
\end{align}

\noindent Although this result was obtained from hydordynamics, note that conformal symmetry requires $\langle C \rangle(t) = \lambda^2(t) \langle C \rangle(0)$. This leads to the following differential equation:

\begin{align}
\frac{d^2\lambda^2(t)}{dt^2} = 2\left(\omega_i^2 + \omega_f^2 \right) - 4\omega_f^2 \lambda^2(t)
\label{eq:lambda_diff}
\end{align}

\noindent Given the initial conditions:

\begin{align}
\lambda(0) &=1 &  \dot{\lambda}(0) &=0,
\end{align}

\noindent the solution to Eq.~(\ref{eq:lambda_diff}) is Eq.~(\ref{eq:lambda_quench}). That is both hydrodynamic and conformal arguments give the same prediction for the dynamical scale factor.

A similar analysis for $t>t_h$ gives the following differential equation:

\begin{align}
\frac{d^2 \lambda^2(t)}{dt^2} &= 2\left(\omega_i^2 + \omega_f^2\right) + 2\left(\omega_i^2-\omega_f^2\right) \lambda^2(t_h) - 4 \omega_i^2 \lambda^2(t)
\label{eq:lambda_diff2}
\end{align}

\noindent where $\lambda(t_h)$ is the value of Eq.~(\ref{eq:lambda_quench}) at $t_h$.

Given the solution for $\lambda^2(t)$ and its first derivative are continuous functions of $t$, the general solution to Eq.~(\ref{eq:lambda_quench}) has the form:

\begin{align}
\lambda^2(t>t_h) &= \frac{E_f}{2\omega_i^2 \langle C \rangle(0)} \nonumber \\
&+ A \cos\left(2 \omega_i(t-t_h)\right) + B \sin\left(2 \omega_i(t-t_h)\right)
\end{align}

\noindent where $E_f$ is the final energy:

\begin{align}
E_f &= \left[(\omega_i^2 + \omega_f^2) +(\omega_i^2 - \omega_f^2) \lambda^2(t_h)\right] \langle C \rangle(0)
\end{align}

\noindent and the coefficients $A$ and $B$ are found to be:
\begin{align}
A &= \lambda^2(t_h) - \frac{E_f}{2\omega_i^2 \langle C \rangle(0)} \nonumber \\
B &= \frac{1}{2\omega_i}\frac{d}{dt} \lambda^2(t_h)
\end{align}

In general the system will exhibit undamped oscillations around $E_f/(2\omega_i^2 \langle C \rangle(0)$ at frequency $2\omega_i$ as required by conformal symmetry. The amplitude of those oscillations is determined via:

\begin{equation}
\text{Ampl.} = \sqrt{A^2 + B^2}.
\label{eq:amp_final}
\end{equation}

\noindent Since $A$ and $B$ are periodic functions with frequency $2\omega_f$, the amplitude of these oscillations can be tuned continuously. Specifically for $t_h = n\pi /\omega_f$, $\lambda^2(t_h) = 1$ and $d\lambda^2(t_h)/dt =0$. In this case, $A = B = 0$ while $E_f = 2\omega_i \langle C \rangle(0)$. Thus the amplitude of the oscillations is zero, and $\lambda(t>t_h) = 1$. This is expected as there is a quantum revival of the full many-body state for this choice of $t_h$.  When $t_h = (2n+1)\pi/2\omega_f$ we have maximum amplitude oscillations with $\text{Ampl.} = \omega_i^2/\omega_f^2$. For arbitrary $t_h$ the amplitude of the oscillations changes continuously between the aforementioned limits.

The presence of oscillations for $t>t_h$ are related to the amount of work done on the system by the two-quench protocol. The work done after this two quench protocol is given by:

\begin{equation}
W(t_h)= \langle H_{\omega_i}\rangle (t = t_h^+) - \langle H_{\omega_i}\rangle (t = 0^-),
\label{eq:avg_work_done_def}
\end{equation}

\noindent where $t= 0^-$ and $t_h^+$ mean right before the first quench, and right after the second quench respectively. 

The average work done is necessarily a function of hold time. Since the many-body wave function is completely oscillatory at a frequency $2\omega_f$, the average work done should also be oscillatory. At times $t= n\pi/\omega_f$ for some integer, $n$, the many-body wave function returns to itself, and the average work done will be zero, while it should be maximum at multiples of half the period, i.e. for $t=(2n+1)\pi/2\omega_f$.

The dynamics of the average work done can be illustrated exactly via conservation of energy and conformal symmetry arguments. The initial energy is given by: $E_0 = \langle H_{\omega_i}\rangle (t=0^-)$. Immediately  after the first quench the energy is given by:

\begin{equation}
\langle H_{\omega_f}\rangle (t=0^+) = \left(\omega^2_f - \omega_i^2\right)\langle C \rangle(0) + E_0
\label{eq:E0+}
\end{equation}

\noindent The dynamics after the first quench conserves both energy and entropy, hence $\langle H_{\omega_f}\rangle(0<t<t_h) = \langle H_{\omega_f}\rangle(t=0^+)$. Immediately after the second quench the energy:

\begin{equation}
\langle H_{\omega_i}\rangle(t=t_h^+) = \left(\omega^2_i - \omega^2_f\right) \langle C \rangle(t_h) + E_0
\end{equation}

\noindent The work done for this quench-quench protocol is simply related to the difference between the moment of inertias at time $t_h$ and its initial value:

\begin{align}
W(t_h) &= \left(\omega_i^2 -\omega_f^2\right) \left(\langle C \rangle(t_h) - \langle C \rangle(0)\right) \nonumber \\
&= \left(\omega_i^2 -\omega_f^2\right) \left(\lambda^2(t_h) - 1\right) \langle C \rangle(0)
\label{eq:work_final}
\end{align}

\begin{figure}
\includegraphics[scale=0.5]{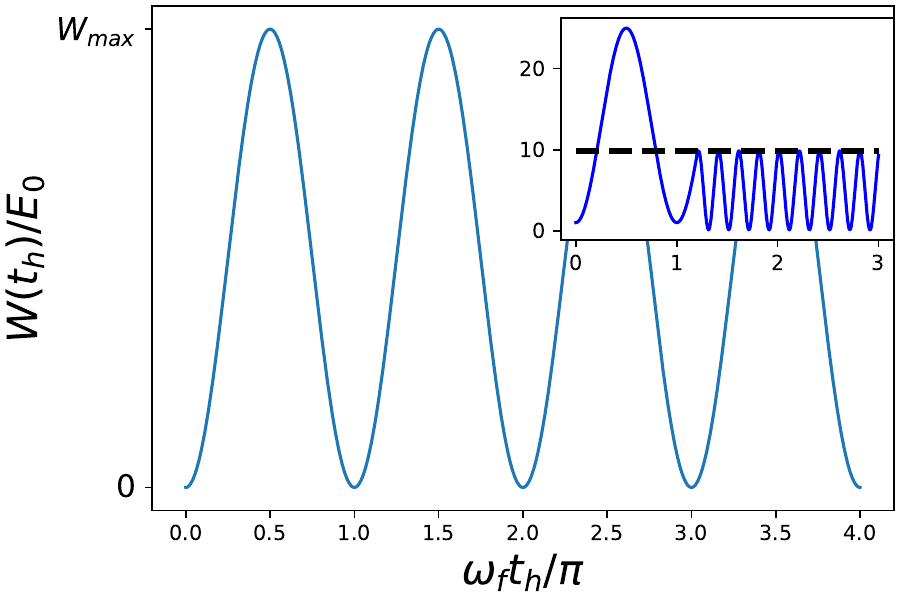}
\caption{Average work done after the two-quench protocol in Eq.~(\ref{eq:omega(t)}) with $\omega_f= \omega_i/5$. The work done is periodic with frequency $2\omega_f$ and oscillates between $0$ and $(\omega_i^2-\omega_f^2)^2/\omega_f^2$. The inset shows the simulation of the time-dependent rescaling factor, $\lambda^2(t)$, during the whole two quench protocol for a generic $t_h \neq n\pi/\omega_f$ for an integer $n$. For $0<t<t_h$, the dynamics are oscillatory with a frequency of $2\omega_f$, while for $t_h < t$, the dynamics are oscillatory with a frequency $2\omega_i$.}
\label{fig:work}
\end{figure}

As expected, the work is an oscillatory function with frequency $2\omega_f$. The work done has a minimum of zero at $t = n \pi/\omega_f$ while it has a maximum value of $W_{max} = (\omega_i^2 - \omega_f^2)(\omega_i^2/\omega_f^2 -1) \langle C \rangle(0)$ at $t= (2n+1)\pi/2\omega_f$. In Fig.~(\ref{fig:work}) we present the average work done from the two-quench protocol according to Eq.~(\ref{eq:work_final}), while the inset shows the residual oscillations of the moment of inertia.

\section{The Zero-Temperature Auto-Correlation Function}
\label{app:auto_corr}

In this appendix we provide a detailed calculation of the zero-temperature auto-correlation function. We first begin with the state-operator correspondence:

\begin{equation}
|\psi_0 \rangle = e^{-i H_{\omega_f}\tau} O_p^{\dagger} | vac \rangle \nonumber
\end{equation}

\noindent where $\tanh(\omega_f \tau) = \omega_f/\omega_i$. The operator $O_p$ is a primary operator, which satisfies the following relations:

\begin{eqnarray}
[C, O_p] =[K_i, O_p]=0.
\label{eq:def_primary}
\end{eqnarray}

\noindent where $C$ and $K_i$ are the generators of conformal transformations and Galilean boosts along the $i = x,y,z$ direction. The scaling dimension of $O$, $\Delta_O$, and the number $N_O$ are given as:
\begin{align}
[D,O_p]&=i\Delta_o O_p & \left[N,O_p^{\dagger}\right] &= N_{O_p} O_p^{\dagger} & \left[N,O_p\right] &= -N_{O_p} O_p
\end{align}

The auto-correlation function is defined in Eq.~(\ref{eq:def_auto_corr}):

\begin{equation}
{G}({\bf R},t) = \left\langle \psi_0 \right| e^{-iH_{\omega_f}t} e^{i {\bf P \cdot R}} \left|\psi_0 \right\rangle \nonumber
\end{equation}

\noindent In the Heisenberg picture, we can write the auto-correlation function as a correlation function of the primary operator:

\begin{equation}
G({\bf R},t) = \langle vac | O_p({\bf R}, t-i \tau) O_p^{\dagger}(0,i \tau) |vac \rangle \nonumber
\end{equation}

\noindent where we have defined:

\begin{equation}
O_p^{(\dagger)}({\bf R},t) = e^{-i{\bf P \cdot R}} e^{i H_{\omega_f}t} O_p^{(\dagger)} e^{-i H_{\omega_f} t} e^{i {\bf P \cdot R}}  \nonumber 
\end{equation}

We can evaluate the auto-correlation function by considering the equations:
\begin{align}
0&= \langle vac | \left[C, O_p({\bf R}, t-i\tau) O_p^{\dagger}(0, i \tau)\right] | vac \rangle \nonumber \\
0&=  \langle vac | \left[K_i, O_p({\bf R}, t-i\tau) O_p^{\dagger}(0, i \tau)\right] | vac \rangle
\label{eq:constraints}
\end{align}
First let us consider:
\begin{align}
\left[C, \right. & \left. O_p({\bf R},t)\right] =e^{-i {\bf P \cdot R}} e^{i H_{\omega_f}t}  \nonumber \\
&\left[ e^{-i H_{\omega_f}t} e^{i {\bf P \cdot R}} C e^{-i {\bf P \cdot R}} e^{i H_{\omega_f}t}, O_P\right] e^{-i H_{\omega_f}t} e^{i {\bf P \cdot R}}
\end{align}

Each of these commutators can be evaluated analytically using the Schroedinger algebra in Eq.~(\ref{tab:commutation_relations}). For simplicity we will report the necessary results below:

\begin{align}
e^{i {\bf P \cdot R}} C e^{-i {\bf P \cdot R}} &= C - {\bf R \cdot K} + \frac{R^2}{2}N \\
e^{i {\bf P \cdot R}} {\bf K} e^{-i {\bf P \cdot R}} &= {\bf K} + {\bf R} N \\
e^{-i H_{\omega_f}t} C e^{i H_{\omega_f}t} &=  \frac{\sin^2(\omega_f t)}{\omega_f^2} H_{\omega_f} - \frac{\sin(2\omega_f t)}{2\omega_f} D \nonumber \\
&+ \cos(2\omega_f t) C \\
e^{-i H_{\omega_f}t} {\bf K} e^{i H_{\omega_f}t} &= {\bf K} \cos(\omega_f t) - \frac{\sin(\omega_f t)}{\omega_f} {\bf P} \\
e^{-i H_{\omega_f}t} {\bf P} e^{i H_{\omega_f}t} &= \cos(\omega_f t) {\bf P}  + \omega_f \sin(\omega_f t){\bf K} \label{eq:Pt}
\end{align}

\noindent We supplement these equations by noting:

\begin{align}
-i\partial_t O_p^{(\dagger)}({\bf R},t) &= e^{-i {\bf P \cdot R}} e^{i H_{\omega_f} t} \nonumber \\
&\cdot \left[H_{\omega_f}, O_p^{(\dagger)}\right] e^{-i H_{\omega_f} t} e^{i {\bf P \cdot R}} \\
i \nabla_{\bf R} O_p^{(\dagger)}({\bf R},t) &= e^{-i {\bf P \cdot R}} e^{i H_{\omega_f} t} \nonumber \\
& \cdot\left[\cos(\omega_f t){\bf P}, O_p^{(\dagger)}\right] e^{-i H_{\omega_f} t} e^{i {\bf P \cdot R}}
\end{align}

\noindent where we have used Eq.~(\ref{eq:Pt}) and Eq.~(\ref{eq:def_primary}).

Combining all the commutation relations together gives us :

\begin{align}
\left[C, O_p({\bf R},t) \right] &= -i \left[ \frac{\sin^2(\omega_f t)}{\omega_f^2} \partial_t -i \frac{R^2}{2}N_{O_p} \right. \nonumber \\
&\left. + \frac{\sin(2\omega_f t)}{2\omega_f} \left({\bf R \cdot \nabla_R} +\Delta_{O_p}\right) \right] O_p({\bf R},t)
\end{align}

\noindent Following the same arguments we can also obtain:

\begin{align}
\left[{\bf K}, O_p({\bf R},t) \right] &= i \left[ \frac{\tan(\omega_f t)}{\omega_f } \nabla_{\bf R} + \cos(\omega_f t) {\bf R} N_{O_p}\right] O_p({\bf R},t)
\end{align}

\noindent Similar results holds for $O_p^{\dagger}({\bf R},t)$, where we replace $N_{O_p^{\dagger}} = -N_{O_p}$, and $\Delta_{O_p^{\dagger}} = \Delta_{O_p}$.

It is straightforward to combine everything together in Eq.~(\ref{eq:constraints}) to obtain differential equations for the auto-correlation function: Eqs.~(\ref{eq:diff_eqn_auto_correlation}-\ref{eq:diff_eqn_auto_correlation1}). We note that in order to obtain Eqs.~(\ref{eq:diff_eqn_auto_correlation}-\ref{eq:diff_eqn_auto_correlation1}) we have disregarded terms that affect the imaginary time evolution as these only describe the normalization of the auto-correlation function, i.e. $G({\bf 0},0)  = 1$.

\end{document}